# Why the Langevin-Debye theory of molecular polarisation fails in gas phase

M. Michelini, ENEA-Casaccia Research Centre, via Anguillarese 301, 00060 Rome, Italy
maurizio.michelini@casaccia.enea.it

The classical polarization formula of Langevin, which holds in the solid/liquid state, does not satisfy many experimental facts in gas phase, especially in diluted gas mixtures. The new formulation of the molecular polarization in gas phase is obtained on phenomenological grounds analysing the motion that polar molecules undergo under an electric field. It is shown that the polarization amplitude in solids/liquids is limited by the high frequency of the molecule oscillations, which intercept and stop within $10^{-13} \div 10^{-14}$ sec the polarizing rotation due to the external field. Conversely, in gas phase the polarizing motion happens during the molecule time of flight ($10^{-6} \div 10^{-11}$ sec) which is so long that the molecules may become nearly aligned to the field. In this case the twisting moment of the field is weakly contrasted by the moment of inertia or by the gyroscopic moment of single molecules. A quantum mechanical analysis of the couple formation in polar gases is given, as well as an assessment of their fraction at any temperature since polar couples have small lifetime. These aspects were not considered in the 1930s when the dipole measurements in gas phase were firstly carried out. The dipoles calculated with the Langevin-Debye equation often agreed, but in other cases they disagreed with the fixed dipole requirement, as shown by the precise experiments of Zahn and Sanger. The explanation of this departure is given in terms of the concepts introduced by the present theory. This new theory put a revolution in the polarization figures assumed by usual polar molecules. For example, very close to the atmospheric ions the water vapour polarization result $10^5 \div 10^6$ times greater than the classical ones. This allows to explain many unsolved problems of the classical physics, such as: i) vapour nucleation in atmosphere (formation of cloud droplets, even containing pollutant polar molecules $SO_2$, $HNO_3$, $O_3$, $H_2S$, CFC, etc.), ii) the buoyancy force experienced by cloud droplets in the atmosphere, which gave puzzling results in the early experiments of Millikan performed with *water* droplets, whereas oil droplets did not, iii) the unexplained cloud electrification phenomenon, etc.

## 1. INTRODUCTION

The formulation given by Langevin about a century ago states that the molecular polarization induced by an electric field upon a substance depends on the temperature only [1]. According to this formulation there is no difference (provided the temperature is the same) between the molecular polarization in condensed or in gaseous phase.

It was known since the 1930s that some liquid and solid substances showed discrepancies in the calculated dipoles when compared with the dipole measurements in gas phase.

From the theoretical viewpoint a thorough analysis of these discrepancies was done in 1936 by L. Onsager [3] which demonstrated, referring to the Debye's polarization equation, that the internal field factor $4\pi/3$ must be multiplied in liquids and solids by a proper function of the dielectric constants. No attempt was done for gaseous substances, because at that time the behaviour of polar molecules in gas phase was believed to be well known. At the end of the 1940s two scientists, E.O.Rice and E.Teller [2], conceived to verify again for some liquid substances the accuracy of the Langevin formula, trying to analyse the limit of accuracy of that formulation.

An indication of the tendency toward *strong* polarization in rarefied gases may be found in the Clausius-Mosotti equation, according to which the experimental molecular polarization is inversely proportional to the gas density. Equating this experimental polarization to the theoretical Langevin formula, P. Debye established in 1912 a theory which was extensively used in the 1930s to calculate the electric dipole of various substances through the measurements of the dielectric constant in gas phase. The application of this polarization formula in gas phase gave initially good results, calculating fixed dipoles under various conditions [4,5,6]. However in certain cases the calculated dipoles did not agree each other [7]. Besides some dipoles showed an increase in the explored interval of temperature, contrarily to any expectation [8,9,10]. This fact was the signal that something did not work in the theory of the molecular polarization in gas phase. Eucken and Meyer [11] developed a theory of variable dipoles, but their effort did not convince.

In this work we search the reason of this failure by analysing the molecular behaviour under the classical mechanics and, in some case, the quantum mechanics. The molecular polarization in gas phase is formulated describing the specific phenomena which characterise the influence of an external field upon polar molecules with high degree of freedom. This characteristic enables the free polar molecules to contrast the field twisting moment by means of the moment of inertia or by the gyroscopic moment when the rotational energy is sufficiently high. This accurate description makes the calculation of the average molecular polarisation more complex than the simple classical formula (see eq.2.1.c) which considers a unique molecular parameter , i.e. the temperature.

There are several phenomena in gas phase where the simple classical polarisation does not explain the experimental evidence:
1- the nucleation mechanism of the water vapour in atmosphere. It is noteworthy that
   the classical model do not consider the high water dipole (1.85 Debye)
2- the mechanism which allows some pollutant molecules with high dipole [ $SO_2$ =1.67 Debye,
   SO =1.55, $HNO_3$ =2.17, OH=1.67, $O_3$ = 0.53, $NH_3$=1.47, $H_2S$ = 1.1, HCl =1.08, the
   CFC's , etc.] to condense through the process described as gas-to-particle conversion [12]
3- the nature of the forces which allow the fog droplets *to float* within controlled air boxes,
   notwithstanding that water density is one thousand greater than air density. This droplet
   buoyancy is linked to the historical puzzle about the Millikan's experiment, where *oil* droplets
   made possible to measure correctly the electron charge, whereas *water* droplets did not,
4- the cloud electrification phenomena during the rein crisis,
5- the equation of state for polar gases,
6- some phenomena involved in the gaseous dielectrics, the corona effect, etc.

Those who know rational answers to these questions don't need to consider this work. However they are kindly urged to make aware the scientific community of these answers.

Some of the above questions have common features that we shall briefly examine.
*a) The ion induced nucleation.*
The molecular polarisation due to the electric field close to each atmospheric ion is negligible when calculated with the classical polarisation theory : $\mu_r \cong (10^{-5} \div 10^{-6})\mu$. As a consequence the intermolecular polar forces are negligible. This fact prevents the possibility of understanding the nucleation phenomenon of the water vapour in the atmosphere.

The description in specific physical terms of the vapour condensation in atmosphere is not yet clearly known. The experience shows that when the air temperature falls to the saturation level, the scattered vapour molecules around a cloud condensing nucleus (CCN) suddenly collapse producing an embryonic liquid droplet.

Unfortunately, the classical nucleation theory does not consider ions (due to the very low classical polarization induced) as active CCN in atmosphere, in spite of their wide diffusion. According to the classical theory the embryonic droplets are electrically neutral. According to the present polarisation theory in gas phase, half cloud droplets originates from positive ions and half from negative ions. Obviously the second scheme permits us to understand easily and correctly the cloud electric phenomena. On the contrary the classical theory, which always deals with neutral cloud droplets, *needs to explain* the formation of high cloud charges during the rain crisis. As remarked by R.T. Feynman [13], the *ad hoc* theories of charge separation resulted either wrong in sign (as the breaking-drop theory) or unreliable, such as the ingenious theory according to which falling droplets receive a negative charge by fusing with large slow-moving negative ions. In the Feynman's words «this theory looks pretty good, but there are also problems because, after a short time, the supply of large ions would get used up…On the contrary the total charge involved in a thunderstorm is very high».

In 1982 Holland and Castleman [14] denounced a «fundamental inconsistency» in the description of the vapour condensation by means of the classical Thomson's free energy equation for charged droplets, suggesting that «many cluster ions have a more ordered structure than can be accounted for in the classical approach». In other words, the ion-induced nucleation does not happen according to the Thomson's formalism.

The new polarization theory in gas phase gives, within the strong polarization region (SPR) of the ions (whose size goes from 2.6 up to 15 micron), figures which are $10^5 \div 10^6$ times the corresponding classical ones. This means that vapour molecules around the ion are almost aligned, so the dipole-dipole forces may produce a molecular attraction with *mutual collapse*, which triggers (in saturated conditions) the vapour condensation.

This phenomenon occurs for any polar molecule (such as the pollutant ones, previously recalled) mixed to the water vapour, when the dipole is *comparable* to the water dipole (co-condensation).

b) *The droplet buoyancy*

Another issue to be explored is the buoyancy force experienced by the cloud/fog droplets. The origin of this force has been demonstrated in a previous work [15]. In few words, the presence of the (vertical) atmospheric field $E_o(x)$, (which in fair weather shows figures around 120 v/m at sea level and decreases rapidly to 1-2 v/m at the first cold trap altitude $x$ =15-20 km), produces the polarization $\mu_z$ of the water molecules within droplets, which is obviously calculated according to the classical theory, because the medium is liquid. Multiplying $\mu_z$ by the number of molecules in a droplet of radius $a$, one obtains the droplet field whose vertical component takes the value

$$(1.1) \qquad E_z(r,x) \cong - B\,(a/r)^3\,E_o(x)$$

where $r$ is the distance from the centre and B is a constant related to the medium characteristics (for instance, water shows $B \cong 116 \div 120$). This field pertains to all neutral droplets, even to the small fraction which forms generally by fusion of two droplets with opposite charge.

The field $E_z(r,x)$ allows each droplet to hook the upper strongly polarised vapour molecules, which transmit to the droplet an upward force due to the buoyancy of the water vapour (m.w.18) when diluted in the air (m.w.28.9). The calculation shows that the greater is the droplet, the higher is the number of hooked vapour molecules. The equilibrium between the droplet weight and the buoyancy force determines the *sustainable* radius, depending on the vapour density and other parameters [15]. When the buoyancy force cannot further increase, the droplet begins to slowly fall. Let's now put attention to the charged droplets, which have a vertical field

$$(1.2) \qquad E_{z,tot}(r,x) \cong - B\,(a/r)^3\,E_o(x) \pm e\cos\theta\,/\,4\pi\varepsilon_o r^2$$

which shows (according to the sign) two values which differ notably when $E_o(x)$ is small, that is over a certain altitude. The droplet field of eq.(1.2) increases for negative charge, whereas it reduces for positive charge. As a consequence, the charged droplets show a different buoyancy force: those with positive ions have sustainable radii considerably smaller («dwarf» population) than those of the «standard» population with negative ions.

c) *The formation of the cloud charge during the rein crisis*

This problem finds its solution in the two issues previously examined. The first one shows that the greatest part of the cloud droplets (excepting the small fraction of the neutral ones) are charged and consist of two populations: the *dwarf* and the *standard*. When an external agent (for instance a decrease of the air temperature) makes the buoyancy force unable to further sustain the droplets, they begin to slowly fall with a velocity proportional to the square radius (Stokes law). Hence the standard droplets fall *more rapidly* than the dwarf ones.

This produces within the clouds an axial *charge unbalance*, with upper positive centres and lower negative centres (cloud electrification) which gives rise to the usual electric phenomena (lightning).

*c) The equation of state for (di)polar gases*

In a *polar* gas the dipole field $E = \mu/4\pi\varepsilon_o r^3$ tends to align the dipoles of two approaching (or leaving) molecules. As a consequence the molecules experience a mutual *attractive* force and show an increasing velocity $v(r)$ during the approach and a decreasing velocity during the removal, so that $v(r)$ has a minimum in the middle of the mean free path $\lambda$ covered between two collisions. Putting $v_w(r)$ the collision speed on the wall, the average kinetic energy of a polar molecule hitting the wall at a distance $r$ from the last collision is

$$m v_w^2(r)/2 \cong kT - (\mu_r^2/4\pi\varepsilon_o)(1/d^3 - 1/r^3) \qquad (1.3)$$

where $\mu_r$ is the molecular polarisation and $d$ is the minimum distance at the collision.

For instance the typical kinetic energy occuring when $r \cong \lambda$ (i.e. $r \gg d$)

$$m v_w^2/2 \cong kT - \mu_r^2/\pi\varepsilon_o d^3$$

is substantially lower than the molecular energy linked to the thermodynamic temperature.

The pressure of a gas equals in general the energy density of the particles *hitting* the wall. Putting $\underline{v_w}$ the collision speed $v_w(r)$ averaged over the mean free path $\lambda$, in the case of identical particles the pressure takes the form

$$p = \delta_w \underline{v_w}^2/2$$

where $\delta_w$ is the local density close to the wall.

The constancy in space of the molecular flux close to the wall furnishes $\delta_w v_w(r) \cong \delta \underline{v}$, where $\delta$ and $\underline{v}$ are the average parametres of the gas. Substituting in the above equation we finally obtain the generalized equation of state

$$p = \delta \underline{v} \underline{v_w}/2 . \qquad (1.4)$$

When the molecules do not exchange forces between them ($\mu = 0$), the velocity shows a constant value $v_o$ as well as the density ($\delta_w = \delta$). Then eq.(1.3) gives $mv_w^2/2 \cong kT$ and the equation of state takes the simple structure related to the ideal gases

$$p_o = \delta v_o^2/2 = (\delta/m) kT. \qquad (1.5)$$

When the forces between molecules in gas phase are calculated with the classical Langevin formula (which gives very low molecular polarizations $\mu_r$) the «correction» introduced into eq.(1.3) is negligible respect to the ideal gas scheme.

In the present theory the *mutual* polarisation between two approaching (or leaving) molecules is much higher than the classical figure, so the related drawing force cannot be neglected.

Presently the equation of state of a polar gas (neglecting the effect of the polar couples introduced in Sec.4C) must be obtained from the Lennard-Jones-Stockmayer potential through laborious calculations involving the second virial coefficient [1].

## 2. FORMULATION OF THE MOLECULAR POLARIZATION

### 2.A. The classical Langevin-Debye theory

The electric field induces on the molecules an average dipole $\mu_{tot} = \alpha_m E_{loc} + \mu_r$ related to the molecule polarizability $\alpha_m$ and to the average component $\mu_r$ of the permanent dipole $\mu$ along the field direction. The local field $E_{loc} = E(\varepsilon_r + 2)/3$ is defined as the field producing polarization within a medium of relative dielectric constant $\varepsilon_r$ crossed by the external field $E$.

From the experimental standpoint, the average dipole induced in a substance of density $\delta$ and relative dielectric constant $\varepsilon_r$ measured at various temperatures, is given (in SI units) by the Clausius-Mosotti equation

$$\mu_{tot} = (3m/\delta)\, 4\pi\varepsilon_o\, [(\varepsilon_r - 1)/(\varepsilon_r + 2)]\, E_{loc}$$

where $m$ is the molecular mass and $\varepsilon_o = 8,85 \times 10^{-12}$ Farad/m is the vacuum dielectric constant. Note that in the SI system the field due to a point source $q$ is coherently defined $E = q/4\pi\varepsilon_o \varepsilon_r r^2$.
In the following we consider «polar» molecules those for which the effect of polarizability can be neglected, that is $\mu_r \cong \mu$.

From the theoretical point of view, the molecular polarization $\mu_r$ is defined by means of the average rotational interaction energy $\ni$ between the permanent dipole and the local field

$$\ni\; =\; <\boldsymbol{\mu} \bullet E_{loc}>\; =\; \mu E_{loc} < \cos\alpha > = \mu_r E_{loc} \qquad (2.1.a)$$

obtaining

$$\mu_r = <\cos\alpha>\mu. \qquad (2.1.b)$$

Under the conditions of validity of the classical mechanics [1], the calculation of $<\cos\alpha>$ requires to weight $\cos\alpha$ over all orientations with the Boltzmann's factor, thus obtaining

$$\mu_r = \mu\, (\coth x - 1/x) = \mu L(x) \qquad (2.1.c)$$

where $x = \mu E_{loc}/kT$ is the variable of the Langevin function $L(x)$. Expansion of $L(x)$ in power series up to the first term gives the so called classical polarization formula

$$\mu_r \cong \mu x/3 = \mu^2 E_{loc}/3kT \qquad (2.2)$$

which is obviously accurate for usual low fields satisfying the condition $x \ll 1$.
The reliability of the Langevin formula has been studied in liquid phase by E.O.Rice and E.Teller [2] with the aim of calculating the permanent dipole $\mu$ of common polar substances by comparing the theoretical $\mu_r$ with the average dipole experimentally induced by the same field in the same substance

$$\alpha_m + \mu^2/3kT \cong (3\,m/\delta)\, 4\pi\varepsilon_o\, [(\varepsilon_r - 1)/(\varepsilon_r + 2)]. \qquad (2.3)$$

For substances diluted in various non-polar solvents, this calculated $\mu$ agrees in general with the dipole measured with absolute methods, such as the deflection of molecular beams in non-homogeneous electric fields. However in some cases it does not agree.
The discrepancies revealed by eq.(2.3) in liquid and solid substances have been studied in 1936 from a theoretical viewpoint by L.Onsager [3], Nobel prize 1968 for Chemistry. He observed that the Debye's theory, which assumed that eq.(2.3) describes exactly the situation for low density substances such as gases, fails if applied to liquids and solids since the strong interaction between the molecules must be taken into account. For instance, this equation underestimates the dipole for liquid water, whose dielectric constant is accurately known, by a factor about 2.
According to Onsager, the original Mosotti theory of the internal field is not applicable to permanent dipoles since only a part of this field contributes to the dipole orientation. The rather complicated theory shows that the term $\mu^2/3kT$ in eq.(2.3) must be multiplied by the function

$$f = 3\,\varepsilon_r\,(n^2 + 2)/(2\varepsilon_r + n^2)(\varepsilon_r + 2)$$

where $n^2 \cong \varepsilon_\infty$ is the approximated value of the dielectric constant at very high frequency.
In the final discussion, he recognised that this correction factor is valid for liquids and solids, not for gases. In the 1938 Onsager came back to this argument [16] discussing the Langevin polarization factor whenever it approaches unity, as in the gases at high pressure.
This historical view appears necessary to better understand the difficulties introduced in the measurements of $\mu$ by assuming that the classical eq.(2.3) gives accurate molecular polarization in gas phase. In fact in the early 1930s many physicists made measurements of the molecular dipoles, finding discrepancies between the results and sometimes even a break of the undoubted result that molecular dipoles have to be considered fixed. This argument will be examined in Sec.5.B.

**2.B. Fundamentals of molecular polarizationin gas phase**

Our analysis about the inadequacy of the Langevin formula when applied to polar molecules in diluted mixtures is based on the observation that it gives polarization figures which may be underestimated by orders of magnitude. This occurs in particular when a polar gas is rarefied, so the time between two collisions is higher than the time of dipole alignment $t_o \cong 1.71(I/\mu E)^{1/2}$ spent by the twisting moment $M = \mu E \sin\alpha$ of the field (for instance about 300 v/m may be sufficient) to rotate the dipole up to $\alpha = 0$ (alignment) against the molecule moment of inertia.

To verify this strong polarization we introduce (Sec.3.A) in the definition $\mu_r = \ni / E$ the average field interaction energy $\ni = \Omega^2 / 2I$ which depends on the angular momentum $\Omega = \mu E \Delta t <\sin\alpha>$ that the field gives up to the molecule in the period $\Delta t$. This procedure leads to the generalised molecular polarisation

$$\mu_r = \mu^2 E \Delta t^2 <\sin\alpha>^2 / 2I . \qquad (2.4)$$

Substituting the time $\Delta t = t_o$ occurring in rarefied gases, the polarization results $\mu_r \cong \mu (1.71^2/2)<\sin\alpha>^2$. With the aid of eq.(3.8), which gives the quantity $<\sin\alpha>$, one finally has $\mu_r \cong 0.572\mu$, a very high figure when compared to the classical polarization $\mu_r \cong \mu^2 E / 3kT \approx 10^{-7}\mu$ calculated with the same field and temperature.

Why the classical theory, which works well in solids/liquids, does not interpret the experimental results in gas phase ? The principle of superimposition of effects applied to the molecules in solids and liquids shows that the polarising moment $M = \mu E \sin\alpha$ depends on the angle amplitude $\alpha(t)$ that the dipole $\mu$ describes in the complex oscillations due to the strong interactions of the nearest molecules. These oscillations show frequencies $\nu_i$ which characterise the compact molecular structure of the solid/liquid substances. The average molecular polarization has to be calculated over the small time $\Delta t = 1/\nu_i << t_o$, so the angular momentum $\Omega$ that the field gives up to the molecule is small and the above defined polarization results very small, in accord with the classical formula.

When eq.(2.4) is applied in the gas phase, the time $\Delta t$ increases notably as well as the polarization. This correct behaviour is not considered in the classical formula.

In Sec.3 it is shown that $\mu_r$ takes in general the form $\mu_r = \mu^2 E / \varphi$ where the contrasting «energy» $\varphi$ balances the field interaction. In Sec.5 it is formulated the generalised polarization eq.(5.1) by comparison with which one finds $\varphi = 3kT / \Lambda$, where the parameter $\Lambda$ depends on the density and temperature of the substance, so the polarisation becomes $\mu_r = \Lambda \mu^2 E / 3kT$.

For instance in gas phase $\Lambda$ reaches figures up to $10^5 \div 10^6$ and the polarisation approaches 1. However in the solid/liquid state $\Lambda \approx 1$, so the *classical* polarisation $\mu_r \approx \mu^2 E / 3kT$ takes place. This synthetic result is independent of the characteristics of the substance, because any *compact* molecular structure (liquid or solid) exhibits about the same figure of $\Lambda$.

Conversely in the gas phase the molecules can freely rotate, so the polarising motion becomes a definite precession of the dipole/molecule around the fixed field direction.

This situation refers in particular to the polar molecules whose rotational energy usually exceeds the threshold $\mu E$, as described in Sec.4.

The elaboration on phenomenological grounds of a molecular polarization theory in gas phase can be correctly done because quantum mechanics is not involved, as it happens in the Langevin's theory [1]. Quantum mechanics is recalled only in the calculation of the pure rotational infrared frequencies due to the absorption and emission by polar molecules and polar couples.

Resuming, we point out that the polarising motion due to the twisting moment $M = \mu E \sin\alpha$ is in general balanced by two inertial reactions:

- the moment of inertia of the molecule, when the existing rotational energy is low ($I \omega_1^2 / 2 << \mu E$)
- the gyroscopic moment $M_g = d\mathbf{C}/dt$, when the molecular rotational energy is high, as it usually happens in the case of pure polar gases ($I \omega_1^2 / 2 \geq \mu E$).

According to this partition the respective analysis will be carried out in Sec.3 and Sec.4.

In Sec.5 the measurements on pure polar gases which in the 1930s gave data in disagreement with the classical Debye equation are discussed in the light of the new polarization theory.

### 3. MOLECULAR POLARIZATION IN DILUTED GAS MIXTURES

In this paragraph a general procedure is presented to calculate the polarization of polar gases in *diluted* mixtures with non-polar gases whose molecules communicate low rotational energies in the collisions. The interaction of polar molecules between them is presently neglected due to the dilution. This section relates in general to molecules whose average rotation energy is insufficient to produce the gyroscopic effect, i.e. $I \omega_1^2 < 2\mu E$ (see Sec.4.A).

During the flight between two collisions, polar molecules behave as electric dipoles with moment of inertia $I$.

The most rapid action of the local field $E_{loc}$ on polar molecules is the rotation of the dipole $\mu$ due to the twisting moment $M = |\mathbf{\mu} \times \mathbf{E}_{loc}| = \mu E_{loc} \sin\alpha$. For instance, close to an ion, the dipole aligns to the field in a time so small that the molecule displacement is of the order of the molecular size. Since within gases the dielectric constant $\varepsilon_r \cong 1$, then $E_{loc} \cong E$ and the field suffix can be dropped. The field $E$ considered in the following does not change appreciably when the molecule moves along a mean free path in the gas. By consequence, during the time of flight $\tau$ between two collisions, the rotational motion satisfies the equation

$$\mu E \sin\alpha + I(d^2\alpha/dt^2) = 0. \qquad (3.1)$$

Putting $k_o^2 = \mu E / I$, this equation firstly shows the rotational velocity

$$|d\alpha/dt| = 2^{1/2} k_o (\cos\alpha - \cos\theta_o)^{1/2} \qquad (3.2)$$

where $\theta_o$ is the initial orientation of the dipole. When $\alpha = \theta_o$ the rotational velocity is zero and rises up to the maximum when $\alpha = 0$ (the dipole is aligned with the field). The time $t_o$ of alignment

$$t_o = \int_{\theta_o}^{o} d\alpha / 2^{1/2} k_o (\cos\alpha - \cos\theta_o)^{1/2} \qquad (3.3)$$

can be physically reached only if $t_o$ is less than the time $\tau$ between two collisions, which change the dipole orientation.

The time $t_o$ depends on the initial orientation $\theta_o$ through non-elementary tabulated functions. Numerical calculations show that the greater the angle $\theta_o$, the longer the time $t_o$.
Analogously, the solution of the eq.(3.1) requires non-elementary functions.
The choice is between non elementary exact functions or approximate explicit functions.
We try to implement the first choice in the paragraph 3.A, while two approximate explicit formulae are derived in paragraph 3.B.

### 3.A. The direct approach and the «contrasting energy»

An exact formulation of the molecular polarization will be here presented. In the general definition $\mu_r = \ni / E$ (see eq.2.*a*) let's put the average rotational energy $\ni$ of the polarising motion

$$\ni = \Omega^2 / 2I \qquad (3.4)$$

as a function of the molecular angular momentum $\Omega$ that the field gives up, on the average, to the dipole during the time of flight

$$\Omega = \int_o^{\tau} \mu E \sin\alpha(t)\, dt = \mu \tau E \, \underline{\sin\alpha}(\theta_o) \qquad (3.5)$$

where $\underline{\sin\alpha}(\theta_o)$ is the average of $\sin\alpha(t)$ within the time of flight for any given initial orientation $\theta_o$. Substituting in eq.(3.4) one finally obtains the fully accurate expression

$$\mu_r = (\mu^2 \tau^2 / 2I)\, E <\sin\alpha>^2 \qquad (3.6)$$

where $\tau$ is the time of flight within the non-polar gas.

The analytical difficulties are concentrated in the numerical calculation of $<\sin\alpha>$, i.e. the average of $\underline{\sin\alpha}(\theta_o)$ over all molecule orientations. A sufficiently approximate expression of $<\sin\alpha>$ is given in paragraph B, which leads directly to an explicit formulation of $\mu_r$.

Eq.(3.6) has a notable physical insight which can be clarified by comparing with the polarization definition $\mu_r = э/E$, so to obtain

$$(\mu E)^2 = э/(\tau^2/2\,I)<\sin\alpha>^2 = э\,\varphi$$

which shows that the quantity

$$\varphi = 2\,I/\tau^2<\sin\alpha>^2 \qquad (3.7)$$

has the dimensions of an energy (see later) involved in the polarization process. By substitution in eq.(3.6) one gets the more expressive form normally appearing in the polarization formula

$$\mu_r = \mu^2 E/\varphi \qquad (3.7.a)$$

which shows that $\varphi$ is great when the polarization is small (for instance in the classical polarization formula $\varphi \approx 3kT$). Conversely, when the polarization is very high ($\mu_r \approx \mu$) the energy $\varphi$ is a little greater than $\mu E$. The polarization depends in general on this contrasting «energy» involved in the polarising motion.

In liquids and solids, owing to the compact structure, $\varphi$ is high since *several* molecules interact strongly with the polarised molecule. In gas phase the dipole interactions between more distant molecules are obviously weaker and consequently $\varphi$ is small. In the following Sec.4 the nature of $\varphi$ will be better understood. Actually $\varphi$ relates to the molecular moment which contrasts the polarising moment. Eq.(3.6) is valid everywhere, provided that $\tau$ is known.

### 3.B. Approximate solutions

In order to calculate the overall average $<\sin\alpha>$ one has to obtain

$$\underline{\sin\alpha}(\theta_o) = (1/\tau)\int_o^\tau \sin\alpha(t)\,dt$$

For computational purposes it may be sufficient to develop $\sin\alpha(t)$ in power series of $\alpha(t)$ up to the sixth term. To this aim the approximate rotation

$$\alpha(t) \cong \theta_o \cos(0.92\,k_o\,t). \qquad (3.8)$$

can be substituted in the power series.

Integrating the power series over time we obtain $\underline{\sin\alpha}(\theta_o,\beta_o)$ as a function of the initial orientation $\theta_o$ and of the dimensionless parameter

$$\beta_o = 0.92\,k_o\tau = 0.92\,\tau\,(\mu E/I)^{1/2} \qquad (3.9)$$

which has a key role in the polarisation. Averaging over all orientations, we finally obtain

$$<\sin\alpha> \cong (\sin\beta_o/\beta_o)\,[\pi/4 + 0.1625\sin^2\beta_o + 0.0437\sin^4\beta_o]\,. \qquad (3.10)$$

A brief comment about the physical significance of $\beta_o$ is necessary.

This parameter increases when the molecular polarization increases and viceversa.

When the condition of alignment $\alpha(t_o) = 0$ (corresponding to $\beta_o = \pi/2$ in accord with eq.3.8 when $t_o = \tau$) is satisfied, the molecules are strongly polarised. This means that the time of alignment results (through eq.3.9)

$$t_o \approx \pi/1.84 k_o \approx 1.71/k_o. \qquad (3.11)$$

This happens in two situations: i) in the rarefied gases, where the time of flight $\tau$ is so high that $\tau \geq t_o$; ii) in presence of high fields (i.e. high $k_o$) where $t_o$ is so small that $t_o \leq \tau$.

For normally occurring low fields or when the time $\tau$ is (due to the high gas density) much lower than $t_o$, we have $\beta_o << 1$.

A reasonable question arises: why $\beta_o$ does not exceed $\pi/2$ whereas the field $E$ may reach $10^9$ v/m ? The answer is that $\tau$ reduces more and more, because the high field (originating a drawing force) increases the density of the polar molecules (see Part II) thus reducing the mean free path $\lambda = \tau/(2kT/m)^{1/2}$ and the time of flight $\tau$.

In *rarefied* gases $\tau$ may reach figures greater than the natural period of the molecule oscillation $\tau_o = \pi/2k_o$ under the polarising field. In this case we must put $\tau = \tau_o = 1.71/k_o$ which gives in turn $\beta_o \cong \pi/2$. In general it is very useful to express eq.(3.6) by means of the parameter $\beta_o^2 = 0.846\, \tau^2 \mu E/I$. After substituting $<\sin\alpha>^2$ by means of eq. (3.10), one has the explicit polarization formula for computational purposes

$$\mu_r \cong 0.5907\mu \sin^2\beta_o\, [\pi/4 + 0.1625\sin^2\beta_o + 0.0437\sin^4\beta_o]^2 \qquad (3.12)$$

which depends only on the parameter $\beta_o = 0.92\,\tau\,(\mu E/I)^{1/2}$ whose range of existence is confined between 0 and $\pi/2$. This relationship gives, without restrictions on the field magnitude, the polarization of molecules scattered in a non-polar gas with time of flight $\tau = \lambda\,(m/2kT)^{1/2}$, which contains both the density and the temperature of the gas.

The most relevant result of this formula is the highest polarisation figure ($\mu_r \cong 0.581\mu$) which can be obtained (contrarily to the classical polarization formula) even with fields of modest magnitude, provided that the time of flight $\tau$ is sufficiently high that $\beta_o = 0.92\,\tau\,(\mu E/I)^{1/2}$ approaches $\pi/2$.
For instance, let's evaluate the time $\tau$ which is necessary to obtain in the atmosphere a *strong* vapour polarization in presence of moderate fields (for instance 300-400 volt/m close to an ion charge, see Sec.6). The required time is lower than the time of flight $\tau \approx 10^{-10}$ at the sea-level, so the *strong* polarisation can be reached everywhere in atmosphere. As a consequence, the water vapour nucleation can be always induced at altitudes lower than the first cold trap (16-20 km).

Resuming, eq.(3.6) gives a reliable formulation of the molecular polarization due to its direct derivation. In the computational form of eq.(3.12), we found that it is also acceptably accurate because the assumptions made in computing $<\sin\alpha>$ produce inaccuracies that can be known.

## 4. MOLECULAR POLARIZATION THEORY IN PURE POLAR GASES

In case of pure polar gases each molecule undergoes the sum of the fields of all surrounding dipoles, which determines an oscillating angular velocity. The rotational energy of polar molecules undergoes something like an equipartition due to the dipole-dipole interaction. Thanks to this considerable rotational energy (which is in general greater than the energy $\mu E$), polar molecules give rise to a precession motion, whose gyroscopic moment balances the field polarising moment.

### 4.A. The polarization of single polar molecules

The polarizing motion of a molecule in a polar gas is described by the same kind of equation used in the preceding paragraph.

Now the polarizing field $E$ acts on a molecule which is always rotating because the sum of the fields $E_{tot} = \Sigma_j E_j(z)$ from all other dipoles generates the twisting moment $M_{tot} = \mu \times E_{tot}$ which influences the rotational energy of the molecule. The dipole interaction links the angular velocities of all polar molecules.

Since the rotational energy is high, the molecule undergoes a dipole precession having the axis in the field direction. To study this motion we recall that there is a linkage between the direction of the electric dipole and the *axis of symmetry* of the molecule *internal* masses. These ones coincide with the charged nuclei and determine the moment of inertia of a linear polar molecule.

In fact the positive dipolar-charge ($+\Delta e$), as well as the negative polar-charge, is placed on the axis of symmetry of the charged nuclei. Hence the dipole direction coincides with the symmetry axis and is orthogonal to the principal axis of inertia of the molecule.

This is the reason why the twisting moment of the field $M = \mu E \sin\alpha$ is balanced by the moment of the centrifugal forces acting on the internal masses $M_{cf} = I\omega_x^2 \cos\alpha \sin\alpha$, where $I$ is the principal moment of inertia. The angular velocity $\omega_x$ of these masses around the central precession axis arises from the composition of the initial molecule rotation $\omega_1$ and of the precession rotation $\omega_p = M / I \omega_1$. Then the equilibrium of the twisting moments in the precession motion is

$$\mu E \sin\alpha = I \omega_x^2 \cos\alpha \sin\alpha \qquad (4.1.a)$$

which shows that $\mu_r = \mu \cos\alpha$ is linked to

$$\mu_r = \mu^2 E / I \omega_x^2. \qquad (4.2)$$

Moreover, the so-called contrasting «energy» $\varphi = I \omega_x^2$ relates to the gyroscopic contrasting moment (which has the dimensions of energy) at the right side of eq.(4.1.a).

The polarization depends in general on the rotation $\omega_x$ which can be obtained from the conservation of energy

$$(I \omega_x^2 / 2) \sin^2\alpha + \mu E \cos\alpha = I \omega_1^2 / 2. \qquad (4.1.b)$$

Eliminating $\omega_x^2$ between eqs.(4.1.a) and (4.1.b) one finds the equation

$$\cos^2\alpha - (I \omega_1^2 / \mu E) \cos\alpha + 1 = 0 \qquad (4.2.a)$$

which, putting $b = I \omega_1^2 / \mu E$, gives real solutions

$$\mu_r / \mu = \cos\alpha = \tfrac{1}{2} [b - (b^2 - 4)^{1/2}]$$

when $b \geq 2$, i.e. when the rotational energy satisfies the condition

$$I \omega_1^2 / 2 \geq \mu E. \qquad (4.2.b)$$

This relationship expresses the well known fact that the gyroscopic effect takes place when the initial kinetic energy is high enough. Polar molecules are effective micro gyroscopes because their rotational energy is well shared between them.

Each angle $\alpha$ which is a solution of eq.(4.2.a) corresponds to a stable precession.

The polarization figures appear to go from the minimum $\mu_r = \mu^2 E / I \omega_1^2$ (when $\mu E \ll I \omega_1^2$) up to the maximum $\mu_r = \mu$ when the field is very high (i.e. $\mu E = I \omega_1^2 / 2$).

On this basis, the contrasting moment for the usual low field conditions is

$$\varphi_1 \approx I \omega_1^2. \qquad (4.2.c)$$

The calculation of the average angular velocity $\varpi_1$, which depends on the molecular translational energy $kT$, needs to study the dipole-dipole interaction.

### 4.B. The interaction between single polar molecules

In the polar gases the dipole-dipole interaction softly transforms the translational into rotational energy and viceversa.

The principle of conservation of the angular momentum of two approaching polar molecules (one of these is considered at rest) rotating in the same plane, can be easily applied choosing the initial condition when the dipoles (after a rather complex rotation) become parallel, so the drawing force is $f = 6\mu^2 \cos^2\gamma / 4\pi\varepsilon_o z^4$, where $\gamma$ is the angle between the dipole direction and the distance $z$.

In these conditions the angular velocity of both dipoles may be put equal to $\omega_0$ and the distance between the two molecules does not generally exceed a fraction of the average distance $z_o$, so the sum of the fields of the nearest molecules (placed beyond the distance $z_o$) acts in the same way on both dipoles and may be neglected. Under these assumptions the principle states that

$$2 I \omega_0 + m v_o z_{om} \cong 2 I \omega_1 + m v_1 z_m. \qquad (4.3)$$

where the right side represents the angular momentum at the time when the distance is minimum ($z_m$) and $\omega_1$ is the maximum angular velocity achieved by *both* molecules, since the interacting dipoles align (i.e. $\gamma = 0$) when their distance is minimum (syncronization).

In addition we have to consider the conservation of energy

$$I \omega_0^2 + m v_o^2 / 2 + 2\mu^2 / 4\pi\varepsilon_o z_m^3 \cong I \omega_1^2 + m v_1^2 / 2. \qquad (4.3.a)$$

The difference in eq.(4.3)

$$m (v_o z_{om} - v_1 z_m) = m v_o \Delta z$$

can be calculated through the transverse component of the dipole drawing force $f$, so the small quantity $\Delta z$ equals $2\mu^2 / 2kT \, 4\pi\varepsilon_o z_m^2$. On the other hand from eq.(4.3) we have the difference

$$(\omega_1 - \omega_0) = (m v_o \Delta z / 2I)$$

which can be compared with eq.(4.3.a) written in the form

$$(\omega_1^2 - \omega_o^2) = m(v_o^2 - v_1^2)/2I + 2\mu^2/I\,4\pi\varepsilon_o z_m^3.$$

where the term $m(v_o^2 - v_1^2) = -\theta_o \mu^2/4\pi\varepsilon_o z_m^3$ is calculated through the increase of translational energy due to the component of the dipole drawing force along the trajectory.

The angle $\theta_o$ (which does not exceeds $\pi/2$) is the initial value of the angle $\alpha$ swept by the moving molecule around the fixed one ($\alpha = 0$ when the distance is minimum).

From the last two equations it follows that

$$\omega_1 + \omega_o = (1 - \theta_o/4)\,2\,v_o/z_m \qquad (4.4)$$

In virtue of the synchronisation the maximum angular velocity of any trajectory is

$$\omega_1 = v_1/z_m \qquad (4.5)$$

Substituting $v_o/z_m \approx \omega_1$ in eq.(4.4) gives the initial angular velocity $\omega_o \cong \omega_1(1 - \theta_o/2)$.

This result has some conceptual content, since indicates that a very soft interaction ($z_m$ large, $\theta_o$ little) does not alter the trajectory, so the ratio ($\omega_o/\omega_1$) remains near 1. When $z_m$ is small, the corresponding interaction bends the trajectory, so $\theta_o$ tends to $\pi/2$ and the ratio ($\omega_o/\omega_1$) becomes small. The interaction does not give rise to a polar couple as far as $z_m > 2r_c$, where $2r_c$ is the diameter of polar couples, as shown in the following paragraph.

The maximum angular velocity achieved by polar molecules is $\omega_1^2 \approx 2kT/4mr_c^2$. For instance, water exhibits $\omega_1 \approx 1.1 \times 10^{12}$ at 0°C, assuming the couple diameter (see eq.4.7.c) $2r_c = 4.48$ Å.

To calculate the average rotational energy $\varpi_1^2$ of single polar molecules for polarization purposes, we must take into account for any trajectory the initial and the maximum angular velocity (eq.4.5), obtaining approximately $<\omega^2> \approx 0.16\,\omega_1^2$.

The average between all trajectories (identified by the distance $z_m$) can be calculated from eq.(4.5) assuming that $z_m$ goes uniformly from $2r_c$ up to about $0.2\,z_o$, thus obtaining $\varpi_1^2 \approx 1.6kT/m\,z_o\,r_c$.

Substituting the distances $z_o$ and $r_c$ one finds the average contrasting energy (as defined by eq.4.2.c)

$$I\varpi_1^2 \approx 3.2\,(I/m)\,(kT)^{4/3}\,(4\pi\varepsilon_o \delta/m\mu^2)^{1/3} \qquad (4.5.a)$$

which is an increasing function of $(kT)^{4/3}$ and of $\delta^{1/3}$.

At ordinary temperature and density, $I\varpi_1^2$ is greater than $2\mu E$ (condition 4.2.b), so the polar molecules work as micro gyroscopes. However in the rarefied polar gases $I\varpi_1^2$ may be less.

Polar molecules at ordinary temperature are mostly arranged as couples in pure polar gases (as later shown), so the single molecules prevalently interact with polar couples.

The relevant difference with the above single-to-single interaction is that the couple rotational energy is much higher than the rotational energy of single molecules, so the angular velocity $\omega_1(t)$ of the approaching molecule rises up to the couple rotation $\omega_c$ when the distance is minimum (syncronization). As later shown, the water couple exhibits $\omega_c \cong 2.7 \times 10^{12}$.

In conclusion the average rotational energy $I\varpi_1^2$ of single-to-single molecules (eq.4.5.a) has to be multiplied by the factor $(\omega_c/\omega_1)^2 \approx 6$ to take into account the high presence of polar couples.

### 4.C. The formation of polar couples and their polarization

Between two parallel dipoles at a distance $z$ there is a drawing force $f(z) = 6\mu^2 \cos^2\gamma/4\pi\varepsilon_o z^4$, where $\gamma$ if the angle between the dipoles and the their joining line.

In the case of polar molecules the classical equipartition of energy ($kT/2$ for each degree of freedom) holds approximately since the molecular velocity varies during the flight. Then the *average* translational energy of polar molecules equals approximately $3kT/2$. While a non-polar couple shows an equilibrium rotational energy $kT$, the rotational energy of a polar couple equals approximately $kT + W_\mu$, where the energy $W_\mu \approx 2\mu^2/4\pi\varepsilon_o(2r_c)^3$ is due to the dipolar drawing force. Within these assumptions the couple shows an equilibrium rotational energy

$$\tfrac{1}{2}I_c\omega_c^2 \approx kT + 2\mu^2/4\pi\varepsilon_o(2r_c)^3 \qquad (4.6)$$

where $I_c = 2mr^2$ is the moment of inertia of the couple. From the standpoint of classical mechanics this couple is not a stable structure (unless the two molecules become a dimer), because the centrifugal force $m\omega_c^2 r$ prevails on the drawing force $f(2r_c)$ at the radius $(r_c + \Delta r)$, while the contrary is true at the radius $(r_c - \Delta r)$. This structure must be analysed from the quantum mechanical viewpoint. The radius of the couple may be calculated requiring that the quantized energy levels $\ni_{cj} = j(j+1)\hbar^2/4m r_j^2$ equal the rotational energy levels of the eq.(4.6), thus obtaining an equation for the diameter

$$m kT_j (2r_j)^3 + 2m\mu^2/4\pi\varepsilon_o = (2r_j) j(j+1)\hbar^2. \qquad (4.6.a)$$

The solution of this equation takes place (following a simple procedure suggested by Feynman /15/) when the energy of the system is minimum, i.e. when the derivative of the left side equals the derivative of the right side. This gives

$$(2r_{jo})^2 = j_o(j_o+1)\hbar^2 / 3m kT_{jo} \qquad (4.7)$$

where $j_o$ is the minimum quantum number corresponding to real solutions. This quantum number

$$j_o(j_o+1) = 3m(\mu^2/4\pi\varepsilon_o)^{2/3} (kT)^{1/3}/\hbar^2$$

is generally high. For instance, $j_o$ is about 78 for $H_2O$ and $HCl$ at 0°C and takes greater figures for heavier polar molecules.

Substituting in eq.(4.6.a), the couple diameter related to the equilibrium temperature takes the value

$$2r_c = (\mu^2/4\pi\varepsilon_o kT)^{1/3} \qquad (4.7.a)$$

which is generally higher than the dimer distance for strong polar molecules at usual temperature.
Introducing the diameter $2r_c$ in the eq.(4.6), the rotational energy becomes

$$\tfrac{1}{2} I_c \omega_c^2 \approx 3 kT. \qquad (4.6.b)$$

Polar couples are not dimers, but only temporary associations of polar molecules which undergo disrupting collisions and continuously form.

The calculated lifetime of vapour couples diluted in atmosphere is of the order of $5\times10^{-10}$ sec. Due to this short lifetime, the most attended quantum number of the couple is linked to the angular momentum inherited in the formation. Conversely in *pure* polar gases the lifetime is higher and the couple quantum number takes the equilibrium figure $j_o(j_o+1)$.

The preceding results show that the quantum standpoint is *essential* because in this frame the dipole drawing force $f(z)$ gives rise to a *stable* equilibrium around the quantized radii.

As eq.(4.7.a) shows, the equilibrium diameter of the couples reduces when the gas temperature increases. At ordinary temperatures the couple diameter exceeds generally the dimer intermolecular distance $d_2$. For instance, water couples at 0°C show a diameter of 4.48 Å, well over $d_2 \approx 3$Å. However large molecules, with about the same dipole, show greater intermolecular distances, so at ordinary temperature they form dimers.

From eq.(4.6.a) the couples show an average square angular velocity

$$\omega_c^2 = 6 kT / 2 m r_c^2 = (12/m)(kT)^{5/3}(4\pi\varepsilon_o/\mu^2)^{2/3} \qquad (4.8)$$

which sensibly increases with the temperature. For instance, water at 0°C shows $\omega_c \cong 2.7\times10^{12}$.

Polar couples predominate when the gas density is low. At constant density the couple concentration becomes higher when the temperature is not far from the critical temperature $T_{cr}$ (see next paragraph).

In the couple polarization process the average contrasting moment $\varphi = I_c \omega_c^2$ is involved.

The field twisting moment $M$ induces on the couple a precession motion during which the joining distance $z$ makes with the rotation axis (i.e. the field direction) a constant angle $\alpha$. Each molecule rotates at a distance $r_c \sin\alpha$ from the axis, with angular velocity $\omega_x$ which results from combining the precession and the initial couple rotation. In these conditions the balance of forces shows an equilibrium between the dipole drawing force and the component of the centrifugal forces along the

joining distance
$$6\mu^2 \cos^2\gamma / 4\pi\varepsilon_o (2r_c)^4 = m\,\omega_x^2\,r_c \sin^2\alpha .$$
Recalling eq.(4.7.a) and the moment of inertia of the couple, one has
$$6kT \cos^2\gamma = I_c\,\omega_x^2 \sin^2\alpha . \qquad (4.9.a)$$
Moreover, the field acts on each dipole making them rotate of a small angle γ, so the twisting moment becomes $M = \mu E \sin(\alpha-\gamma)$. Between the two parallel centrifugal forces there is a distance $(2r_c \cos\alpha)$ which originates the contrasting moment $M_c = 2m\omega_x^2 r_c^2 \sin\alpha \cos\alpha = I_c\,\omega_x^2 \sin\alpha \cos\alpha$.
Hence the balance of the moments is
$$\mu E \sin(\alpha-\gamma) = I_c\,\omega_x^2 \sin\alpha \cos\alpha . \qquad (4.9.b)$$
The conservation of energy requires, recalling also eq.(4.6.a),
$$(I_c\,\omega_x^2/2)\sin^2\alpha + \mu E \cos\alpha = I_c\,\omega_c^2/2 \cong 3kT . \qquad (4.9.c)$$
Since in the usual applications $\mu E \approx (10^{-4} \div 10^{-7})\,kT$ can be neglected, this equation reduces to $I_c\,\omega_x^2 \sin^2\alpha \cong 6kT$ which substituted in eq.(4.9.a) gives $\cos\gamma \cong 1$, $(\gamma \cong 0)$.
Substituting these expressions in the eq.(4.9.b) one gets $\mu E \sin^2\alpha = 6kT \cos\alpha$.
Then $\cos\alpha$ is the solution of the equation
$$\cos^2\alpha + c\,\cos\alpha - 1 = 0$$
where $c = 6kT/\mu E$. Since for usual applications $c \gg 1$, then $\cos\alpha \cong (1/c)$.
The molecular polarization of the couple, defined in general as $\mu_r = \mu \cos(\alpha-\gamma)$, in the present conditions becomes $\mu_r \cong \mu \cos\alpha$, which usually equals
$$\mu_r \cong (\mu/c) \cong \mu^2 E / 6kT. \qquad (4.10)$$
This gives for the polar couples a contrasting energy $\varphi = I_c\,\omega_c^2 \cong 6kT$.

### 4.D. The fraction of couples in pure polar gases

The couples undergo destroying collisions between them, but continuously form by collapse of single molecules. It is questionable whether the collisions with single molecules can disconnect the couples whose binding energy is 3 times $kT$. It is also questionable that in a collision between them both couples are destroyed. Under these assumptions the balance of the couple concentration $n_c$ is
$$dn_c/dt = n_s/2t_{mc} - \sigma_c\,n_c\,\phi_c \qquad (4.11)$$
where $t_{mc}$ is the time of mutual collapse of a pair, $\sigma_c \approx g\,(2r_c)^2$ is the couple disruption cross section (the factor $g$ is probably around unity), $\phi_c = v_c\,n_c$ is the flux of couples, where the translational velocity is $v_c = (kT/m)^{1/2}$.
As shown in Sec.8 the molecules originated from a couple disruption remove (under the dipole-dipole drawing force) up to a distance $z$ and begin to collapse in the time given by eq.(8.2) $t_{mc} = z^{5/2}\,(12\pi\varepsilon_o\,m)^{1/2}/5\mu$. The sum of the total molecules remains obviously constant $(\delta/m) = 2n_c + n_s$. Then in stationary conditions eq.(4.11) becomes
$$n_s = 2t_{mc}\,\sigma_c\,n_c^2\,v_c . \qquad (4.12)$$
Defining the fraction $x_s = 1 - 2x_c$ and substituting in the above equation, we obtain
$$x_c^2 + x_c\,(m/v_c\,t_{mc}\,\sigma_c\,\delta) - m/2\,v_c\,t_{mc}\,\sigma_c\,\delta = 0$$
which gives the couple fraction
$$x_c = a\,[(1 + 1/a)^{1/2} - 1]. \qquad (4.12.a)$$
where the parameter $a = m/2\,v_c\,t_{mc}\,\sigma_c\,\delta$, becomes
$$a = (1.443/g)\,(m/\delta)^{1/6}\mu/(4\pi\varepsilon_o\,kT)^{1/2}(2r_c)^2$$
by substitution of the above expressions for $\sigma_c$ and $t_{mc}$, where the distance $z$ is equal to the average molecular distance. From eq.(4.7.a) it appears that the couple diameter $2r_c$ depends on the temperature, although it cannot be lower than the dimer intermolecular distance $d_2$.
The diameter equals $d_2$ when the temperature exceeds the critical figure
$$T_{cr} = \mu^2/4\pi\varepsilon_o\,kd_2^3 \qquad (4.13)$$
whereas $2r_c$ is given by eq.(4.7.a) at lower temperatures.
The relevant property of $T_{cr}$ is that the function $a(T)$ presents a cuspidal point at $T = T_{cr}$, so that

$$a \approx (m/\delta)^{1/6} \mu / d_2^2 (4\pi\varepsilon_o kT)^{1/2} \qquad \text{when } T \geq T_{cr} \qquad (4.13.a)$$
$$a \approx (m/\delta)^{1/6} (4\pi\varepsilon_o kT)^{1/6} / \mu^{1/3} \qquad \text{when } T < T_{cr}. \qquad (4.13.b)$$

As a consequence the parameter $a(T)$ attains the maximum when $T = T_{cr}$.
The couple fraction $x_c$ increases when $a$ increases and takes the maximum ($x_c = \frac{1}{2}$) when $a >> 1$.
For instance, non-saturated water vapour at 0°C shows the figure $a \approx 8.2$ and by consequence
$x_c \approx 0.485$, that is about 97% of the molecules are arranged as couples.
The critical temperature, calculated for some usual small polar molecules ($H_2O$, $NH_3$, $OH$, $HCl$, etc.), shows figures well over the ambient temperature, due to the little distance $d_2$. For instance, water shows a $T_{cr} \approx 800°K$. However, the distance $d_2$ of large molecules is sensibly greater than 3Å of water, so their $T_{cr}$ may become lower than the ambient temperature.
This was the case of the ethylene chloride, a molecule with about 100 molecular weight.
The measurements upon specimens of this gas at constant density [9,10] gave in the 1930's a dipole *increasing* with the temperature, contrarily to any expectation. Since the dipole was calculated by means of the Debye's equation, this departure cast a doubt on that polarization theory.
This experimental result relates to the fact that ethylene chloride shows a $T_{cr}$ (236°K) lower than the explored temperatures, so the parameter $a$ (see eq.4.13.a) reduced when the temperatures increased. In turn, also the couple fraction reduces, incrementing by consequence the fraction of single molecules. This change in the molecular aggregation fully interprets the experimental data, producing a fixed dipole through the explored interval of temperatures, as detailed in Sec.4.C.

## 5. COMPARISON BETWEEN CLASSICAL AND NEW POLARIZATION FORMULA

Eq.(3.6) may induce some scepticism comparing its structure with the classical formula.
However the simple substitution $\tau^2 = \lambda^2 (m/2kT)$ gives the following expressive definition of the molecular polarization in diluted gas mixing

$$\mu_r \cong b(\beta_o) [\mu^2 E / 3kT] (m\lambda^2/I). \qquad (5.1)$$

The classical polarization in square brackets is multiplied by the new factor
$$\Lambda = b(\beta_o)(m\lambda^2/I) \qquad (5.2)$$
which combines the moment of inertia with the m.f.p. in diluted gas mixing.
The coefficient $b(\beta_o) = \frac{3}{4} <\sin\alpha>^2$, defined by means of eq.(3.8), varies slightly from $b \cong 0.462$ (far from the SPR) up to $b \cong 0.299$ (within the SPR).
Let's note that in the new polarization equation (as in the original eq.3.6) no practical limitation of the field is required, contrary to the classical formula.
Incidentally we note that comparing eq.(5.1) with eq.(3.7) the contrasting energy $\varphi$ receives a definite formulation $\varphi = 3kT/\Lambda$, which shows that in gas phase (where $\Lambda$ attains high figures, as shown in the following paragraph) $\varphi$ may be much lower than the classical $3kT$.
To give an idea of the numerical polarization figures, we do some calculation relative to the water molecule, whose moments of inertia depend on the presence of two Hydrogen atoms at a distance of 0.967 Å from the Oxygen nucleus. The maximum $I_{max} = 2.96 \times 10^{-47}$ appears when the H nuclei rotate in the molecule plane. The minimum $I_{min} = 1.05 \times 10^{-47}$ when the two nuclei are specular respect to the plane of rotation. In the following we assume the average value $I \cong 2 \times 10^{-47}$.
As regards the vapour nucleation problem, we have to compare the two theories at distances of 1-2 microns from the ion, that is within the SPR, where the parameter $\beta_o = \pi/2$. According to the new molecular polarization given by eq.(3.6.a) $\mu_r \cong 0{,}581\mu$, whereas the classical polarization formula gives in the above range of distances $\mu_r = (2 \div 8) \times 10^{-7} \mu$. This very low figure explains why the water dipole did not give in the past any contribution to the ion nucleation theory.
We may also deduce that, when the vapour density tends to that of the liquid (as it happens in the condensation process), eq.(5.1) tends to the classical polarization formula which holds for liquids. In fact $\Lambda$ tends to figures around unity in agreement with the well known discrepancies in the liquid state discussed by Rice&Teller [2] and Onsager [3].

### 5.A. Vapour polarization and cloud buoyancy in atmosphere

The parameter $\Lambda$ is important in the cloud buoyancy theory [14], according to which the highly polarized vapour molecules, (which receive an upward push from the heavier air molecules), sustain the weight of droplets whose field is able to capture them. In fact substituting in eq.(5.2) the vapour m.f.p. $\lambda_{eq} = v_o \tau_{eq}$ calculated according to eq.(5.4), one obtains $\Lambda \approx 10^5$.

This high value of $\Lambda$ has a simple physical explanation. The polarization of vapour molecules (see eq.3.6) increases with the time of flight $\tau$, during which the field $E$ produces the dipole rotation in a quite undisturbed manner. When the gas is rarefied, the time of flight $\tau$ may exceed the natural period of the free polar molecule oscillation $t_o \cong 1.71/(\mu E/I)^{1/2}$, so $t_o$ must be substituted in eq.(3.6) to calculate the polarization.

In the troposphere the time of flight $\tau$ within the air varies from $1.6 \times 10^{-10}$ (sea level) up to $1.5 \times 10^{-9}$ sec. The natural period $t_o$ of oscillation due to the atmospheric field (which in fair weather conditions is about 120 volt/m at sea level and about 2 volt/m at 15-16 km) varies from $2.8 \times 10^{-10}$ up to $1.9 \times 10^{-9}$ sec, so it is a little greater.

In general the calculation of the polarization requires to take into account the contribution of the polar gases. The interaction between single polar molecules (Sec.4), although their fraction in atmosphere is low at ordinary temperatures, produces a notable molecular polarization $\mu_r = \mu^2 E / I \omega_1^2$ (see eq.4.2.b).

Polar molecules are mostly bound in temporary couples with an angular velocity $\omega_c$. This corresponds to a high contrasting energy $\varphi_c \approx 6kT$ (see eq.4.10).

As a general rule, the total polarization $\mu_r$ of a molecule (with dipole $\mu$ and polarizability $\alpha$) diluted in a gas mixing is defined by summing up the three contributions, so the polarization of the dielectric medium results

$$P = (\delta/m) \mu_r = \Sigma_i (\delta_i /m_i) (\alpha + \mu^2/\varphi_i) E \qquad (5.3)$$

where the contrasting energy is for single molecules $\varphi_1 = I \omega_1^2$, whereas for couples is $\varphi_2 \approx 6kT$. The contribution of the non-polar molecules depends on the equivalent contrasting energy (eq.3.7) $\varphi_{np} = 2 I / \tau^2 <\sin\alpha>^2$.

It may be sometimes useful to know the equivalent time period, which approximately is

$$\tau^2_{eq} \approx \Sigma_i (\delta_i /m_i) (I/\varphi_i) / \Sigma_i (\delta_i /m_i). \qquad (5.4)$$

This figure may be substituted in the definition of $\beta_o (\tau_{eq})$ in order to calculate the molecular polarization $\mu_r (\beta_o)$ through eq.(3.6.a). This procedure is appropriate to the calculation of the polarization in the rarefied gases.

### 5.B. Dipole measurements using the Debye's polarization formula

In the measurements of the dielectric constant in gas phase it is necessary to know the magnitude of the field $E$ which is opportune to use. From the definition of the dipole field $E_1(z) = 3\mu \cos\gamma \sin\gamma / 4\pi\varepsilon_o z^3$, in the case of two approaching polar molecules at a distance $z$ the disturbing moment $\mathbf{M_1} = \boldsymbol{\mu} \times \mathbf{E_1}(z)$ equals the polarizing moment $\mathbf{M} = \boldsymbol{\mu} \times \mathbf{E}$ when $E_1(z) \approx E$, that is when the distance satisfies the relationship $z_x^3 \approx 3\mu \cos\gamma \sin\gamma / 4\pi\varepsilon_o E$.

When $z \leq z_x$ the interacting molecules drop the achieved polarization. To reduce this disturbance the external field has to be much greater than the field $E_1(z_x)$. Since $z_x$ is of the order of the average molecular distance $z_o$, it is necessary that the experimental field must obey the condition

$$E >> (3\mu \delta / 2m \, 4\pi\varepsilon_o). \qquad (5.5)$$

In the case of water vapour, the experimental field must be $E >> 3 \times 10^6 \delta$. This condition may be difficult to reach in the gas polarization experiments, needing fields as high as $E \approx 10^6$ v/m.

In general the measurements are performed at usual fields and vapour densities, so that polar molecules sensibly influence each other and give also rise to temporary couples.

In the early 1930s there was a great interest among the physicists to measure the dipole of various molecules in gas phase by adopting eq.(2.3)

$$4\pi\varepsilon_o (\varepsilon_r - 1)/(\varepsilon_r + 2) = (\delta/3m)(\alpha_m + \mu^2/3kT) \qquad (5.6)$$

which was considered exact in the frame of the Debye's polarization theory (1912) assuming that polar molecules *do not influence each other*.

In 1924 C.T.Zahn [4] developed a heterodyne null method to measure in gases at atmospheric pressure the dipoles of some hydrogen-alogen compounds: HCl, HBr, HI up to 300°C.

The figures he obtained through the Debye's equation agreed with the theory of fixed dipoles, but they were about two times those arising from a Pauli's quantum theory.

Moreover, the measurement from infra-red absorption of HCl gave results 6 times higher than the Debye's values.

In 1926 Zhan [5] measured the polarization of water vapour at ambient temperature and variable pressure from 3 to 20 mmHg. He found that the polarization showed an abnormal change of the slope against pressure as long as the temperature did not exceed 47°C.

Successively, in 1928 J.D. Stranathan [6] made measurements by dilute solutions in non-polar liquids on the methyl, ethyl, propyl and amyl alcohols, which had shown discrepancies against the Debye's theory when measured as pure liquids. Dielectric constants were obtained at various temperatures, from freezing to boiling. Using the zero concentration intercepts for each substance, leaded to dipole figures substantially independent of the temperature.

In the same year Zahn and J.B.Miles [7] measured the dipoles of CO, COS, $CS_2$, $H_2S$ and did a comparison of all measurements made by others with the same apparatus, thus finding notable discrepancies in the data.

In 1930 C.T.Zahn [8] made measurements in gas phase of the acetic acid dielectric constant at various temperatures and pressures, which indicated a departure from the Debye's equation since the calculated dipole showed an increment of 20% in the temperature interval.

The anomaly was discussed in terms of quantum mechanics, giving an explanation in terms of a transition to a higher vibration state of the OH group.

One year later C.T.Zahn [9] tried to measure in gas phase the dipole of the ethylene chloride in the temperature range from 32°C to 270°C. These observations showed «a definite departure from the Debye's theory» since the dipole plotted against the temperature showed an increase of about 36%.

Zahn made an accurate analysis of this anomalous behavior examining the effect of the molecular association and other possible effects. He compared also with Sanger's values [10] obtained with the same method over a smaller range of temperature. These data appeared less variable, but the corresponding polarizability intercepted on the ordinate axis resulted almost twice that of the optical refractivity. The situation appeared to be obscure.

In the Nobel lecture he pronounced in 1936, P. Debye recalled the «precise measurements» published by C.T. Zahn in 1924, but did not mention or discuss the adverse results that Sanger and Zahn published in 1930-'32. This is not a comment, is a record. A possible reason is that the eminent physicist, as well as others scientists which analysed the experimental data, didn't know the explanation of the experimental departure. Even now many physicists do not suspect that the classical polarization formula is incorrect in gas phase.

### 5.C. Dipole measurements using the new molecular polarization

In the case of pure polar gases (Sec.4) a molecular polarization formula (eq.4.2) was found for single molecules. As the polar molecules easily associate to form couples, the relative polarization formula is given by eq.(4.10).

Now we have to verify whether, using the contrasting energy $\varphi_1 = I\varpi_1^2$ for single molecules and $\varphi_2 = 6\,kT$ for couples, it is possible to correctly calculate the dipole of a pure polar gas from data collected in the previously mentioned polarization measurements.

Substituting the total theoretical polarization $\mu_{tot} = \alpha_m E_{loc} + \mu_r$ in the Clausius-Mosotti formulation we obtain the correct equation to be used in deriving the dipole figures from the measurements, assuming that $x_c$ percent of the gas particles works as couples and $(1-2x_c)$ percent works as single molecules

$$4\pi\varepsilon_o (\varepsilon_r - 1)/(\varepsilon_r + 2) = (\delta/3m)[\alpha_m + \mu^2(x_c/6kT + (1-2x_c)/I\omega_1^2)]. \qquad (5.7)$$

To compare the classical equation (5.6) with the general eq.(5.7) in the assumption that each theory possibly gives a different dipole figure ($\mu_o$ for classical, $\mu$ for present theory), one must equal the molecular polarization arising from each theory to the measured values

$$\mu^2[x_c/6kT + (1-2x_c)/I\omega_1^2] = \mu_o^2/3kT = [\mu_r/E]_{exp}. \qquad (5.8)$$

Both theories must interpret the polarization measurements giving a fixed dipole.

The fraction $x_c$ could be taken from eq.(4.12.a) which predicts figures sufficiently sound. However the numerous assumptions made in deriving that equation advise to follow a more reliable way based on the experimental data.

As previously said, the measurements made between 1924 and 1933 often showed agreement, but in some definite cases showed discrepancies with the classical polarization theory [8,9,10,11]. In particular the measurements relative to the ethylene chloride [9,10,11] showed a molecular polarization $[\mu_r/E]_{exp}$ nearly constant in the temperature range from 300°K up to 550°K, so the dipole $\mu_o$ calculated assuming the Langevin-Debye equation was necessarily increasing with the temperature, differently from the structure of the molecules. The experimental data by Zhan [9] showed that $\mu_o(T)$ increased of about 36% in the temperature range from 305°K up to 543°K.

Incidentally we note that the ratio $\mu_o(T)/\mu$ (assuming the presently known value $\mu = 6.9 \times 10^{-30}$) was about 0.75 at 543°K. Conversely, Sanger's data [10] show that the present value was reached at about 370°K. However his curve did not intercept the correct polarizability. The situation was extremely complex.

When these data are interpreted by means of eq.(5.7) we observe that the calculated dipole does not vary with temperature if one assumes a particular function $x_c(T)$ which satisfies the measured polarization. Recalling eq.(4.5.a) we obtain the ratio y between the two contrasting energies

$$y = I\omega_1^2/6kT \approx 3.2(I/m)(kT)^{1/3}(4\pi\varepsilon_o\delta/m\mu^2)^{1/3} = C_i(kT)^{1/3}. \qquad (5.9)$$

Since the measurements were made at constant gas density, thus $C_i$ is constant for each experiment and by consequence y increases with the temperature.

Introducing y in eq.(5.8) and putting the measured molecular polarization $(\mu_r/E)_{exp}$ equal to a constant B, one gets the fraction of couples

$$x_c(T) = (1 - 6kT B y/\mu^2)/(2 - y). \qquad (5.10)$$

With reference to the specific case of the ethylene chloride, we calculate from the experimental data $B \approx 1.2 \times 10^{-39}$ and $y \approx 9 \times 10^5 (kT)^{1/3}$. Substituting in eq.(5.10) we finally obtain the couple fraction in the temperature range 305÷543°K

$$x_c(T) \approx [1 - 1.35 \times 10^{26}(kT)^{4/3}]/[2 - 9 \times 10^5(kT)^{1/3}]. \qquad (5.11)$$

This relationship shows the reduction of the couple fraction when the temperature increases. So it makes possible to evaluate the change of the fractions of couples and of single molecules when the experimental temperature increases from 305°K up to 543°K. The couple fraction reduces from 0.4898 up to 0.4386 (-10%), whereas the fraction of single molecules increases from 2.04% up to 12.3% (6 times). These data reasonably explain why the experimental molecular polarization remained constant in the explored temperature interval.

The failure of the classical theory to interpret the experimental data of ethylene chloride depend on the fact that this large polar molecule has a critical temperature (as defined by eq.4.13) $T_{cr} = 236°K$ which is *lower* than the experienced temperatures. By consequence, the couple fraction decreases with the temperature, as shown in Sec.4.D.

We have also to discuss why most of the measurements gave fixed dipoles using the Langevin-

Debye's equation. This happens when the couple fraction increases with temperature, that is for the molecules whose critical temperature $T_{cr}$ is well over the experienced temperatures, (such as the light polar molecules $H_2O$, $HCl$, , $NH_3$ , etc).

The argument of the molecular temporary association was several times advanced in works of the 1930s. However it could not solve the problem since the Debye's theory ignored that the molecular polarization in gas phase depends even on the rotational energy of the single molecules $I\varpi_1^2$.

To assess this matter it would be very desirable an experimental device to determine at any temperature/density the fraction of single molecules (or couples) in a polar gas.